\documentclass[twocolumn,aps,prl,amsmath,floatfix]{revtex4}
\usepackage{graphicx}
\begin{document}
\title{Correlation induced spin freezing transition in FeSe: 
       a dynamical mean field study} 
\author{Ansgar Liebsch$^1$ and Hiroshi Ishida$^2$}
\affiliation{$^1$Institut f\"ur Festk\"orperforschung, 
             Forschungszentrum J\"ulich, 
             52425 J\"ulich, Germany\\ 
             $^2$College of Humanities and Sciences, Nihon University,~Tokyo 156,
             Japan}  
\begin{abstract}
The effect of local Coulomb interactions on the electronic properties 
of FeSe is explored within dynamical mean field theory combined with 
finite-temperature exact diagonalization.
The low-energy scattering rate is shown to exhibit non-Fermi-liquid 
behavior caused by the formation of local moments. Fermi-liquid properties
are restored at large electron doping. In contrast, FeAsLaO is shown to 
be located on the Fermi-liquid side of this spin freezing transition. 
\\
\mbox{\hskip1cm}  
electron systems 
\end{abstract}
\maketitle

The recent discovery of high-temperature superconductivity in iron-based 
pnictides~\cite{kamihira} and chalcogenides~\cite{hsu} has led to an intense 
discussion concerning the role of Coulomb correlations in these materials. 
Although compounds such as FeAsLaO (1111), BaFe$_2$As$_2$ (122), LiFeAs (111), 
and FeSe (11) all have rather similar one-electron properties, 
a variety of experiments suggest significant differences. 
For instance, photoemission measurements show that FeAsLaO is moderately 
correlated, with about 50\,\% $3d$ band narrowing and effective mass 
enhancement of about 2 to 3~\cite{maleb}.  
In contrast, several photoemission data on FeSe$_x$Te$_{1-x}$ samples reveal 
larger effective mass enhancement and stronger band narrowing
~\cite{yoshida,yamasaki,xia,nakayama,tamai}, optical measurements exhibit
incoherent spectral features indicative of a pseudogap~\cite{chen}, 
and several transport measurements show large deviations from 
Fermi-liquid behavior~\cite{pallechi,tropeano,song,scales}.  
Moreover, recent theoretical work~\cite{miyake} suggests that Coulomb 
interactions in (11) compounds ought to be less well screened
than in (1111) systems.          
   
The aim of this paper is to elucidate the origin of the experimentally
observed bad-metallic behavior of FeSe. Using an accurate single-particle
description of the electronic structure together with appropriate
interaction parameters~\cite{miyake}, and evaluating the influence of local 
Coulomb interactions within dynamical mean field theory (DMFT)~\cite{dmft}, 
we show that correlations are strong enough to give rise to
the formation of Fe $3d$ local moments, implying non-Fermi-liquid behavior, 
where electronic states at the Fermi energy exhibit a finite lifetime. 
Moreover, we demonstrate that these properties are caused by a nearby 
doping-driven spin-freezing transition, i.e., Fermi-liquid behavior is 
restored towards larger electron doping, whereas hole doping reinforces 
bad-metallic properties. 
Using the same approach for FeAsLaO we show that Coulomb interactions 
are too weak to cause spin freezing, so that this system merely exhibits 
moderate effective mass enhancement~\cite{aichhorn1}.

Thus, FeSe appears to be a material which exhibits bad-metallicity induced
by a spin freezing transition, a mechanism recently identify by Werner 
{\it et al.}~\cite{werner} in a three-band model. For FeAsLaO, Haule 
{\it et al.}~\cite{haule} showed that large Coulomb interactions lead 
to the formation of local moments and large scattering rates,
while Ishida and Liebsch~\cite{prb2010} discussed the spin freezing 
transition as a function of Coulomb interaction and doping. 
Bad-metallic behavior in FeSe was also found in DMFT studies by Craco 
{\it et al.}~\cite{craco} and Aichhorn {\it et al.}~\cite{aichhorn2}. 
The spin-freezing origin of this behavior, however, was not investigated
in these works. 

To account for local Coulomb interactions among Fe $3d$ electrons we use 
exact diagonalization~\cite{ed} (ED) which we have recently extended 
to five orbitals~\cite{prb2010}. Discretization of the lattice surrounding 
the local impurity is achieved by using ten bath levels, which yields 
excellent projections of the lattice Green's function onto 
the cluster consisting of impurity plus bath. Because of the very large 
size of the Hilbert space of this 15 level system (the largest spin sector
has dimension $\sim 40\times 10^6$) the spacing of excited states is very
small so that finite size errors are greatly reduced. Moreover, ED has the 
advantage of allowing for rotationally invariant Hund exchange. 
This is of crucial importance for the spin freezing transition since 
omission of spin-flip and pair-exchange interactions leads to a 
significant shift of the Fermi-liquid to non-Fermi-liquid phase boundary. 
For computational reasons we restrict ourselves to the lowest excited
states, which are relevant near the $T\rightarrow 0$ limit. 
Only the paramagnetic phase is considered. 
Further calculational details can be found in Ref.~\cite{prb2010}. 

The electronic properties of FeSe are formulated 
in terms of the effective low-energy model recently derived by Miyake 
{\it et al.}~\cite{miyake}. In this scheme, standard band structure
calculations within the local density approximation (LDA) were carried 
out, analogous to previous work in Refs.~\cite{subedi,lee}.  
From these results, maximally localized Wannier functions were constructed 
for the Fe $3d$ bands, following the procedure discussed 
in Ref.~\cite{souza}. The inter-site transfer integrals
are then derived from the matrix elements of the Kohn-Sham Hamiltonian
$H({\bf k})$  in the basis of these Wannier functions. Finally, the 
constrained random-phase-approximation (RPA)
developed by Aryasetiawan {\it et al.}~\cite{ferdi} was used to 
determine the effective Coulomb and exchange interaction parameters,
which account for screening via Se $4p$ orbitals. Because of the 
planar structure of FeSe, these screening processes affect the various $3d$ 
orbitals differently so that the interaction parameters exhibit 
appreciable orbital dependence. 
According to these results, the key feature of FeSe is that Coulomb 
interactions are typically 50\,\% larger than in FeAsLaO. This follows
from the larger $z$ spacing and concomitant smaller spread of Wannier 
orbitals, and from the reduced number of screening channels in FeSe. 
The one-electron transfer integrals $t_{mn}({\bf R})$ and interaction 
matrices $U_{mn}$ and $J_{mn}$ for FeSe are given in Tables VII and VIII
of Ref.~\cite{miyake}, respectively. The average intra-orbital Coulomb 
interaction is $\langle U\rangle=4.2$~eV and the average Hund exchange
is $\langle J\rangle=0.5$~eV. The $x,y$ axes point along 
Fe second-neighbor directions.

\begin{figure} [t!] 
\begin{center}
\includegraphics[width=4.5cm,height=6.5cm,angle=-90]{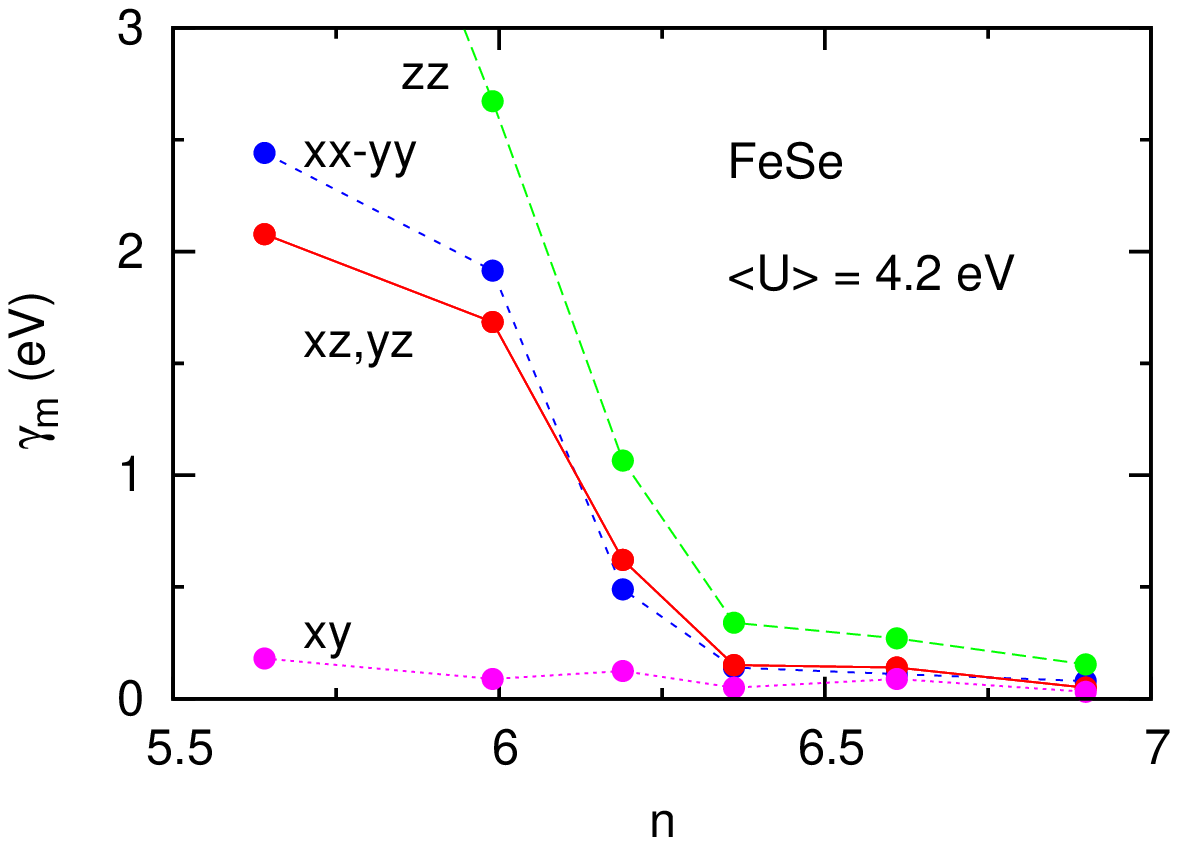}
\end{center}
\vskip-7mm \ \ \ (a)\hfill  \mbox{\hskip5mm}
\begin{center}
\vskip-5mm
\includegraphics[width=4.5cm,height=6.5cm,angle=-90]{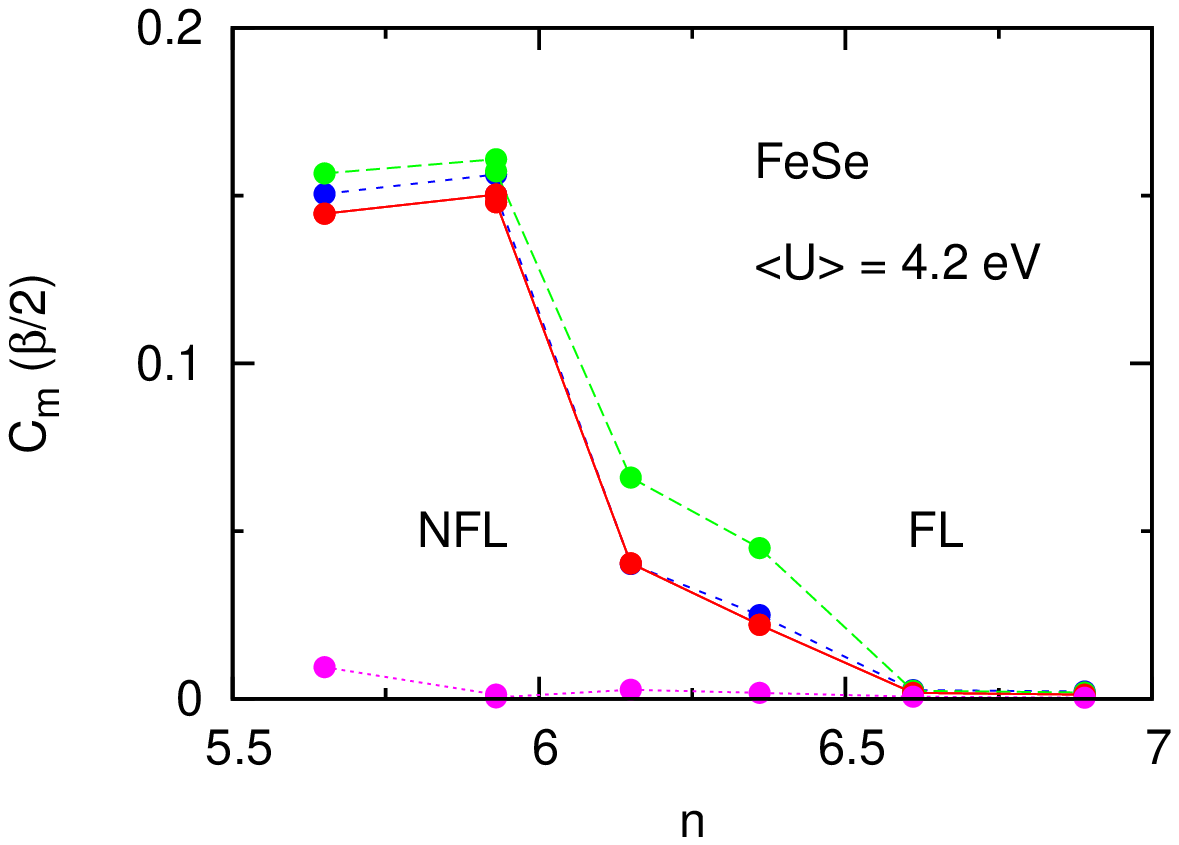}
\end{center}
\vskip-7mm \ \ \ (b)\hfill  \mbox{\hskip5mm}\vskip-1mm
\caption{(Color online)
(a) Low-energy scattering rates and (b) spin-spin correlations 
of FeSe as functions of $3d$ occupancy. Near $n=6$, the 
non-Fermi-liquid (NFL) behavior $\gamma_m>0$ is caused by the formation 
of local moments. Increasing electron doping restores Fermi-liquid (FL) 
properties, whereas hole doping enhances bad-metallic behavior. 
}\end{figure}

Fig.~1(a) shows the low-energy limit of the $3d$ self-energy 
components, $\gamma_m = -{\rm Im}\,\Sigma_m(i\omega_n\rightarrow0)$, 
as a function of Fe $3d$ occupancy. The one-electron transfer
integrals and interaction matrix elements are kept fixed at the values 
for $n=6$. Ordinary Fermi liquid behavior is found for $n>6.5$, but 
increasing low-energy scattering rates are obtained at 
lower electron doping, except for the $d_{xy}$ orbital. 
Even stronger scattering occurs on the hole doping side, $n<6$. 
The reason for the different behavior of $\gamma_{xy}$ is that 
$\rho_{xy}(\omega)$ is about 0.5~eV wider than the other density of 
states components~\cite{miyake}.  

To identify the origin of this change from Fermi-liquid to non-Fermi-liquid 
behavior we have evaluated the spin-spin correlation
function $C_{m}(\tau)=\langle S_{mz}(\tau)S_{mz}(0)\rangle$, where 
$\tau$ denotes imaginary time. In the case of a Fermi-liquid, these
functions decay with $\tau$, so that $C_{m}(\tau=\beta/2)$ is very small
($\beta=1/T=100$~eV$^{-1}$). The susceptibilities 
$\chi_m\sim \int_0^\beta d\tau\, \langle S_{mz}(\tau)S_{mz}(0)\rangle$ 
then are Pauli-like, i.e., independent of temperature.
Fig.~1(b) shows that the results for $n>6.5$ are consistent with this 
behavior. At smaller electron doping, the $C_{m\ne xy}(\tau)$ 
components reach increasingly larger constant values near $\tau=\beta/2$. 
The corresponding susceptibilities $\chi_m$ then become proportional 
to $1/T$, as expected for Curie-Weiss behavior.      

A similar spin-freezing transition was recently found by Werner 
{\it et al.}~\cite{werner} for a degenerate three-band model near $n=2$, 
i.e., at one electron away from half-filling.
Using continuous-time quantum Monte Carlo DMFT, the paramagnetic phase 
at moderate $U$ was shown to exhibit Fermi-liquid properties at small $n$. 
For $n>1.5$, an incoherent metallic phase appears, where the self-energy 
exhibits a finite onset at $\omega=0$ due to the formation of local moments. 
Because of particle-hole symmetry, the same Fermi-liquid to non-Fermi-liquid 
transition appears on the electron-doped side, i.e., when reducing the 
occupancy below $n\approx 4.5$. 

\begin{figure} [t!] 
\begin{center}
\includegraphics[width=4.5cm,height=6.5cm,angle=-90]{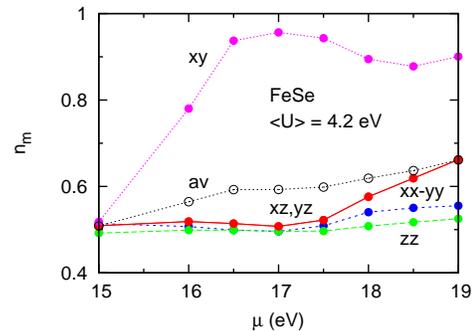}
\end{center}\vskip-3mm
\caption{(Color online)
Fe $3d$ orbital occupancies as functions of chemical potential. 
$n_{av}$ denotes the average occupancy. Nominal occupancy $n=6$ is 
reached near $\mu=17$~eV, while at $\mu=15$~eV
all bands are approximately half-filled.
}\end{figure}

Fig.~2 shows the $3d$ orbital occupancies as a function of doping. At $n=6$, 
these values differ significantly from their uncorrelated values: 
$n_{xz,yz,x^2-y^2,xy,z^2}= (0.54,0.54,0.56,0.60,0.76)$, indicating 
strong interorbital charge transfer induced by Coulomb interactions.      
In particular, $n_{xy}$ reaches nearly unity, while the other $n_{m}$
are close to half-filling. 
Despite this orbital polarization, the spectral distributions 
(not shown here) reveal considerable intensity near $E_F$, suggesting
that the system is not in a Mott phase, where one of the subbands is
filled and the others are split into lower and upper Hubbard bands.
These kinds of partial Mott transitions, with certain subbands empty
or full, are found in several three-band $t_{2g}$ materials, such as 
LaTiO$_3$~\cite{pavarini,LTO}, V$_2$O$_3$~\cite{keller,prb08}, and 
Ca$_2$RuO$_4$~\cite{prl2007,gorelov}.

To investigate the relationship between orbital polarization 
and local moment formation, we have calculated the effect of  
Coulomb correlations in a fully degenerate five-band model consisting 
of identical densities of states for a Bethe lattice, with band width 
and Coulomb interactions of similar magnitude as in FeSe. 
As shown in Fig.~3, this model system exhibits the same kind of 
spin-freezing transition as discussed above for FeSe. For $n=6$, 
the low-energy scattering rate indicates non-Fermi-liquid behavior, 
associated with the formation of local moments. Fermi-liquid behavior 
is recovered at electron doping $n>6.2$, whereas hole doping strengthens 
bad-metallic properties. The important conclusion from this result is that 
orbital polarization is not a prerequisite for the spin-freezing transition.
Thus, this picture differs from the orbital-selective, itinerant-localized 
scenarios proposed in Refs.~\cite{wu,hackl,kou,luca}.

\begin{figure} [t!] 
\begin{center}
\includegraphics[width=4.5cm,height=6.5cm,angle=-90]{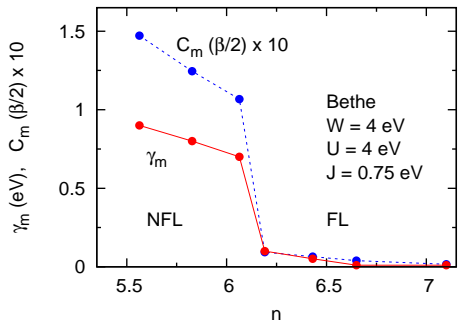}
\end{center}\vskip-3mm
\caption{(Color online)
Low-energy scattering rates and spin-spin correlations 
derived within ED/DMFT for a degenerate five-band model 
with Bethe lattice density of states with $W=4$~eV.
For $U=4$~eV, $J=0.75$~eV, the spin-freezing transition occurs at
$n=6.2$, with Fermi-liquid behavior at $n>6.2$
and increasing bad-metallicity for $n<6.2$. For $U=4$~eV, 
$J=0.5$~eV (not shown), the transition occurs at $n=6.1$.
}\end{figure}

To understand better to what extent the results for FeSe shown in Fig.~1
depend on the details of the single-particle Hamiltonian, we have 
carried out analogous ED/DMFT calculations by replacing $H({\bf k})$ with 
the tight-binding Hamiltonian derived by Graser {\it et al.} for 
FeAsLaO~\cite{graser}. Since the $3d$ density of states components of this 
compound are qualitatively similar to those of FeSe, there ought to be 
also a spin-freezing transition. 
Indeed, if we retain the FeSe interaction parameters with
an average Coulomb energy of $4.2$~eV, the overall behavior of 
the low-energy scattering rates $\gamma_m$ and spin-spin correlations
$C_m$ are qualitatively similar to the ones shown in Fig.~1, except for 
a less pronounced orbital polarization. For instance, 
at $n=6$, $n_{xy}=0.77$ rather than $0.95$. Nevertheless, the 
Fermi-liquid to non-Fermi-liquid transition also occurs near $n=6.5$, 
suggesting that this phenomenon is remarkably robust, i.e., insensitive 
to the details 
of the one-electron properties. These results further support the key 
point of the degenerate-band example presented in Fig.~3, namely, that 
the spin freezing transition is not driven by orbital polarization. 

\begin{figure} [t!] 
\begin{center}
\includegraphics[width=4.5cm,height=6.5cm,angle=-90]{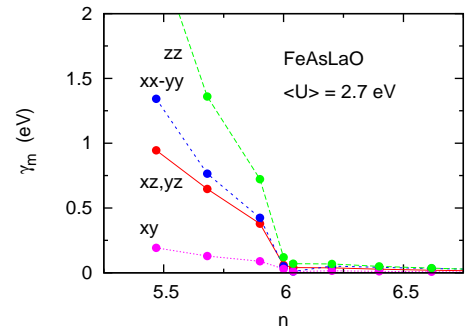}
\end{center}
\vskip-7mm \ \ \ (a)\hfill  \mbox{\hskip5mm}
\begin{center}
\vskip-5mm
\includegraphics[width=4.5cm,height=6.5cm,angle=-90]{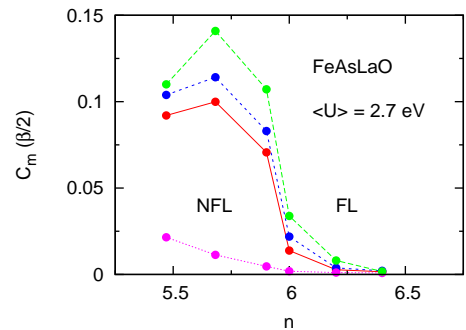}
\end{center}
\vskip-7mm \ \ \ (b)\hfill  \mbox{\hskip5mm}\vskip-1mm
\caption{(Color online)
(a) Low-energy scattering rates and (b) spin-spin correlations 
of FeAsLaO as functions of $3d$ occupancy. The Fermi-liquid to   
non-Fermi-liquid transition caused by local moment formation 
occurs close to $n=6$. Electron doping maintains Fermi-liquid            
properties, whiles hole doping leads to bad-metallic behavior.
}\end{figure}

According to the work by Miyake {\it et al.}~\cite{miyake}, the crucial
difference between FeSe and FeAsLaO is that, because of more efficient $dp$ 
screening and more extended Wannier orbitals, local Coulomb interactions 
in the latter system are considerably smaller, with 
$\langle U\rangle\approx 2.7$~eV and $\langle J\rangle\approx 0.4$~eV. 
In previous ED/DMFT calculations for FeAsLaO~\cite{prb2010} based on 
orbital independent Coulomb and exchange energies we showed that spin 
freezing may occur at about $U=3$~eV for Hund exchange $J=0.75$~eV. 
Using the recently published orbital dependent $U_{mn}$ and $J_{mn}$ matrices 
derived within constrained RPA~\cite{miyake}, we are now able to make a more 
accurate prediction of the spin freezing transition in FeAsLaO. 
The transfer integrals and interaction parameters are again held fixed at 
their values for $n=6$.
As shown in Fig.~4, the transition now is located almost exactly at $n=6$.
Thus, in contrast to FeSe, Fermi-liquid properties in FeAsLaO prevail and
the $3d$ bands exhibit only moderate effective mass enhancement,  
$m^*\approx 2\ldots3$~\cite{aichhorn1}. 
Nonetheless, hole doping $n<6$ gives rise to local 
moment formation and bad-metallic behavior, with a finite lifetime at $E_F$, 
while electron doping stabilizes the Fermi-liquid properties. 
(Note the weaker orbital polarization compared to the results for FeSe in 
Fig.~1. Also, because of orbital dependent $U_{mn}$, $J_{mn}$, the
magnitudes of $\gamma_m$ and $C_m(\beta/2)$ differ from those for orbital 
independent $U, J$ in Ref.~\cite{prb2010}.)   

On the basis of these results we arrive at the phase diagram depicted in 
Fig.~5. FeSe is located well inside the non-Fermi-liquid phase, 
whereas FeAsLaO lies on the Fermi-liquid side of the spin-freezing 
transition. Both systems exhibit `parent' Mott phases in the limit $n=5$.  

\begin{figure} [t!] 
\begin{center}
\includegraphics[width=4.5cm,height=6.5cm,angle=-90]{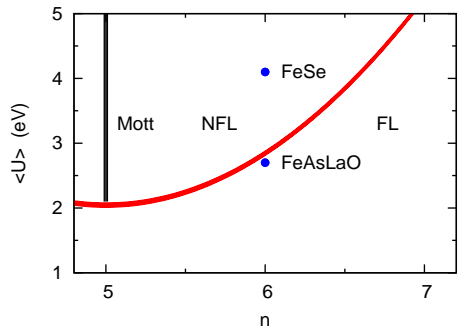}
\end{center}\vskip-3mm
\caption{(Color online)
Schematic phase diagram for FeSe and LaFeAsO. Solid curve: spin freezing 
transition indicating the boundary between Fermi-liquid and non-Fermi-liquid 
phases. Vertical bar at half-filling: Mott insulating phase. 
}\end{figure}
    
We finally discuss the role of spin-flip and pair-exchange interactions.
In Ref.~\cite{prb2010} it was shown that the omission of these terms in the
many-body impurity treatment leads to a significant shift of the Fermi-liquid to 
non-Fermi-liquid phase boundary to $\sim 1$~eV smaller $\langle U \rangle$. 
Thus, for FeSe bad-metallic behavior would become even stronger.
Since FeAsLaO, however, lies very close to the phase boundary, this 
approximation has severe consequences. Instead of a Fermi-liquid 
characterized by moderate effective mass enhancement, this compound is 
then also bad-metallic, in conflict with experiments.

The present work is based on a consistent combination of single-particle
Hamiltonian, constrained RPA interaction parameters, and DMFT
many-electron calculations within a $5\times5$ $d$ electron basis. 
It would be interesting to compare the spin freezing transition obtained
within this scheme to analogous formulations within a $pd$ electron 
basis, such as the one used in Ref.~\cite{aichhorn2}. 
    
In summary, we have evaluated the effect of local correlations in FeSe, 
using accurate single-particle properties and constrained RPA results 
for the orbital dependent Coulomb and exchange matrices, combined with 
ED/DMFT. The finite scattering rates derived from the $3d$ components 
of the self-energy are shown to be linked to the formation of local moments. 
Thus, FeSe is bad-metallic, in agreement with experimental findings. 
Spin freezing ceases for large electron doping, giving rise to Fermi-liquid 
behavior. In contrast, hole doping enhances bad-metallicity. Applying the 
same scheme to FeAsLaO we find that, as a result of the more efficiently 
screened Coulomb interactions, local moment formation is confined to hole 
doping $n<6$. Thus, FeAsLaO and FeSe seem to be located on opposite 
sides of the Fermi-liquid to non-Fermi-liquid spin freezing transition.

A. L. likes to thank M. Aichhorn, R. Arita, E. Gull, K. Haule, A. Millis 
and Ph. Werner, for fruitful discussions. 
The calculations were carried out on the J\"ulich Juropa computer.
This research was supported in part by the National Science Foundation
under Grant PHY05-51164.

\end{document}